\documentclass[aps,twocolumn,epsfig,showpacs]{revtex4-1}
\usepackage{graphicx}
\usepackage{epsfig}
\begin{document}
\title{Statistical Mechanics of DNA unzipping under periodic force: Scaling 
behavior of hysteresis loop}
\author{Sanjay Kumar and Garima Mishra}
\affiliation{Department of Physics, Banaras Hindu University,
     Varanasi 221 005, India} 
\begin{abstract}
A simple model of DNA based on two interacting polymers has been used
to study the unzipping of a double stranded DNA subjected to a periodic 
force.  We propose a dynamical transition, where without changing the 
physiological condition, it is possible to bring  DNA from the 
zipped/unzipped state to a new dynamic (hysteretic) state by varying the 
frequency of the applied force. Our studies
reveal  that the area of the hystersis loop grows with the same exponents 
as of the isotropic spin systems. These exponents are amenable to 
verification in the force spectroscopic experiments.
\end{abstract}
\maketitle

The mechanism
involved in the separation of a double stranded DNA (dsDNA) into two single 
stranded DNA (ssDNA) is a prerequisite for understanding  processes like
replication and transcription. {\it In vitro}, opening of DNA 
is achieved either by increasing the temperature (85-90$^\circ$C) 
termed as thermal melting or by changing the pH value of the solvent
 ($< 3$ or $>9$) called DNA denaturation 
\cite{Wartel}. However, such a drastic change in the physiological condition
is not possible in living systems. The mechanism of opening of dsDNA {\it 
in vivo} is quite complex and is initiated by helicases, DNA and RNA
polymerase, ssb proteins {\it etc}, which exert force of the order of pN and 
as a result DNA unwinds. It is now possible to unzip the two strands
of a DNA using techniques like optical tweezers, atomic force microscopy, 
magnetic tweezers {\it etc} \cite{Bockelmann,prentiss}. Theoretical 
understanding of DNA unzipping is  mostly based on equilibrium conditions 
\cite{bhat99,nelson,Mukamel,maren2k2}. 

However, living systems are open systems and never at equilibrium. 
Understanding the separation of DNA in the equilibrium is 
one approach, but another route is to perform the analysis in 
a situation which closely resembles the living systems {\it i.e.} 
in non-equilibrium conditions. Moreover, helicases are ATP driven 
molecular motors. The periodic hydrolysis of ATP to ADP can generate  
a continuous push and pull kind of motion which spans a wide range 
of length and time scales. As a direct consequence of these 
chemo-mechanical cycles, biological machines act like a 
repetitive force generators, and it is believed that forces with 
periodic signatures are experienced by biomolecules in many physiological
contexts. For instance, it has been postulated that DNA-B, a ring like 
hexameric helicase, pushes through the DNA like a wedge and produces
unidirectional motion and strand separation \cite{dnab}. Active rolling
model and inchworm model are two mechanisms, which suggest
 that PcrA goes through cycle of pulling the ds part of the DNA and then moving 
on the ss part during ATP hydrolysis \cite{pcra}. Williams and Jankowsy 
\cite{nphii} showed that viral RNA helicase NPH-II  hops cyclically 
from the ds to the ss part of DNA and back during the ATP hydrolysis cycle.
Apart from these examples, there are several studies \cite{johnson,
basu,Fili, Szymczak}, which suggest 
that the  force acting on DNA (at the junction of the Y fork i.e. 
ssDNA and dsDNA) 
is periodic in nature rather than constant.  Surprisingly, in most of 
the studies (experiments, theories and simulations), the applied force or 
loading rate is kept constant \cite{kumarphys}, and hence the results 
provide a limited picture of the DNA opening. Application of periodic force 
would introduce new aspects, which are not possible in the steady
force case.

In DNA unzipping, the equilibrium response of the reaction coordinate 
(extension ($y$)) to the constant force is well understood 
in terms of simple models amenable to statistical mechanics 
\cite{kumarphys,marko,kumar_prl}. However, when a dsDNA is driven 
by an oscillatory force, $y$  will also oscillate and lag 
behind the force due to the relaxation delay. This relaxation delay 
induces hysteresis in the force-extension ($f-y$) curve, 
which has been recently observed in simulations and experiments 
\cite{mishra,hatch,kapri}. The nature of hysteresis and its dependence on the
amplitude ($F$) and frequency ($\nu$) of the applied force is well  studied 
in the context of spin systems \cite{madan1,madan2, dd, bkc}. It is found 
that the area under the hysteresis loop $A_{loop}$ scales as 
$F^{\alpha}\nu^{\beta}$.  The values of $\alpha$ and $\beta$ differ from system 
to system \cite{bkc}. However, for DNA unzipping, the non-equilibrium 
response of $y$ to the oscillatory force remains elusive.

\begin{figure}[t]
\includegraphics[width=3.in]{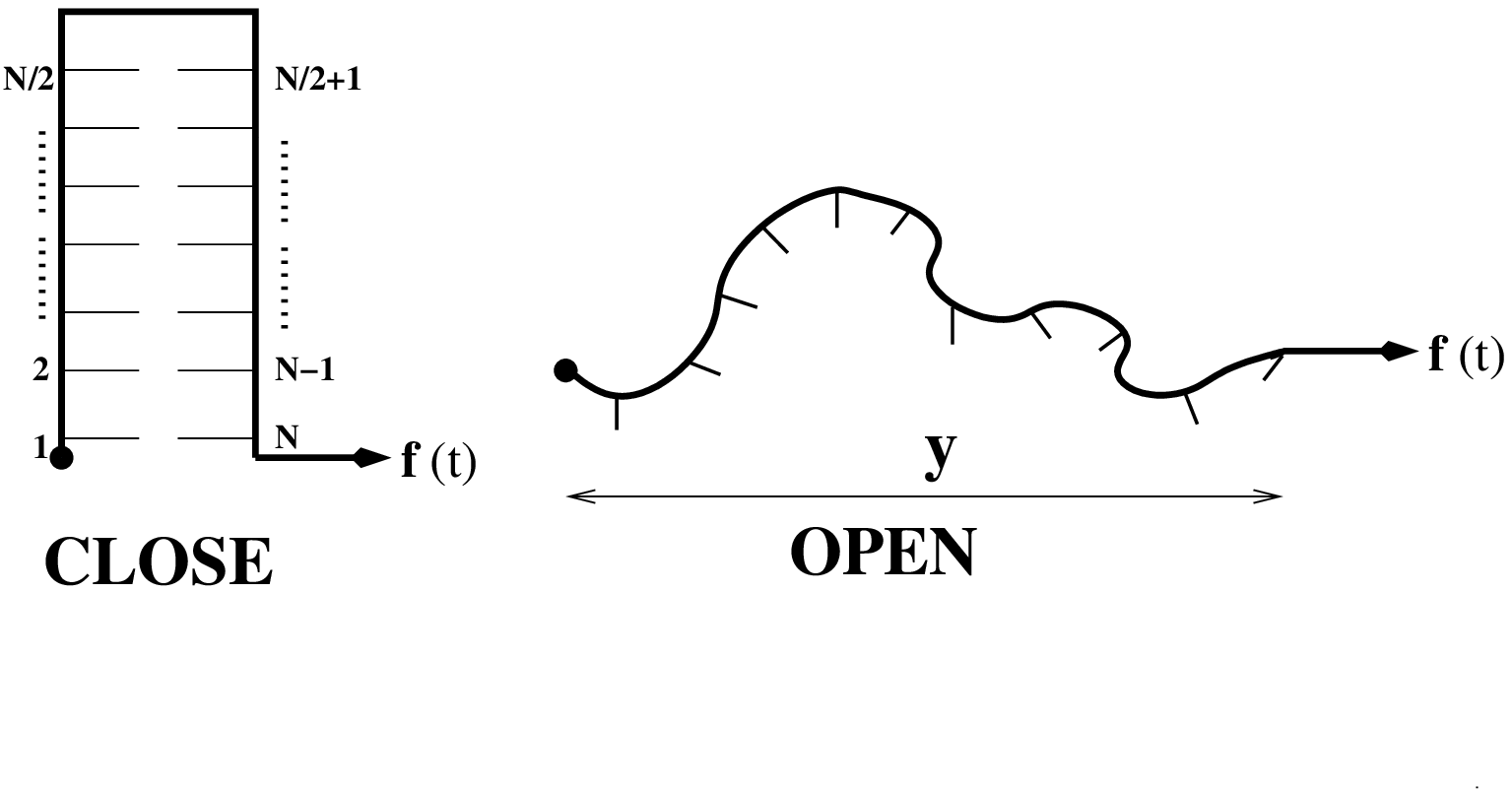}
\caption{DNA in zipped and unzipped state. One end 
is fixed and the other end is subjected to a periodic force.}
\end{figure}
In this Letter, we show that
under a certain physiological condition, a dsDNA remains in the steady and 
stable (zipped or open) state for an extended period of time. Furthermore, 
without any change in temperature ($T$) or pH of the solvent, by varying 
$\nu$ alone, a dsDNA may be brought from the time averaged 
open or  zipped state to a new dynamic state (hysteretic), oscillating
between the zipped and unzipped states, which is dynamical in origin 
and vanishes in the quasi static limit \cite{bkc}. We evaluate the scaling 
exponents $\alpha$ and $\beta$ associated with $A_{loop}$, 
which are amenable to verification in the force spectroscopic experiments. We 
also show that using the work theorem \cite{ps11}, it is possible to 
extract the equilibrium $f-y$ curve from the non-equilibrium pathways.

We consider a simple model of DNA (Fig.1), where a string of beads (complementary
nucleotides) corresponds to a single strand. The beads are connected by
harmonic springs. The excluded volume and and base-pairing interactions
are modeled by the Lennard Jones potential.  It was recently shown that the model
captures some of the essential properties of DNA and the equilibrium
force-temperature diagram is in good agreement with the two state model
in the entire range of the f and T \cite{mishra}. In order to study the dynamical
stability of DNA under the periodic force $f(t)$, we add an energy
$-f (t)y(t)$ to the total energy of the system and perform Langevin Dynamics
simulation to monitor the separation $y$ of the terminal base pairs 
\cite{sm,arxiv}. The random force $\Gamma$ \cite{sm,arxiv}, has also been 
super imposed 
on the periodic force to take account of stochastic fluctuations of the system.
Here, one may fix $\nu$ and vary $F$ or keep $F$ constant and vary $\nu$.  
The value of $f$ increased to its maximum value $F$ in  $m_s$ steps at 
interval  $\Delta f (= 0.01)$ and then it is taken to $0$ in the same 
way \cite{arxiv}. 
Since we are interested in the non-equilibrium regime, we allow only  $n$ 
 LD time steps (much below the equilibrium time) in each increment of 
$\Delta f$. We keep sum of the time spent $\tau (= 2nm_s)$ in  each force 
cycle constant to keep $\nu (=1/\tau)$ constant. In the following, 
we keep $T=0.1$  and $F > 0.32$ \cite{text1,ru}.

\begin{figure}
\includegraphics[width=3.4in]{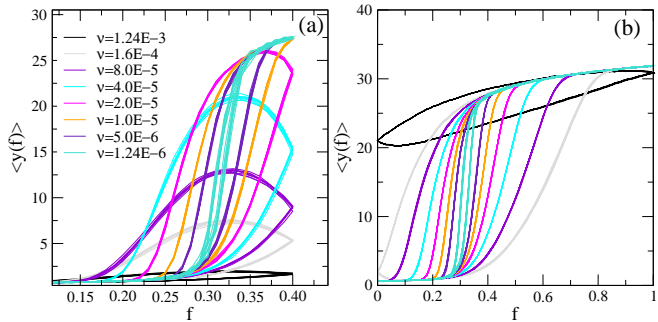}
\caption {$\langle y \rangle$ of DNA as a function of the cyclic force
of amplitudes (a) 0.4 and (b) 1.0 at different $\nu$. 
(a) At high $\nu$, DNA remains in the zipped state with a small hysteresis 
loop. As $\nu$ decreases, the system extends from the zipped state to the 
open state with a bigger loop. For $\nu \rightarrow 0$, the hysteresis 
loop vanishes and the system approaches to the equilibrium path. 
(b) DNA remains in the open state at high $\nu$ and approaches the equilibrium
path from below as $\nu \rightarrow 0$.}

\end{figure}
 
 In Fig.2, we plot the value of $y(f)$ (averaged over C =1000 cycles) 
with $f$ for different values of $\nu$. It is interesting to note that 
the average value of $y(f)$ for different initial conformations 
remains almost the same, showing that the system
is in the steady state \cite{text2}. All plots show hysteresis.
The area under the loop is the  measure of the energy dissipated over a 
cycle and is defined as a dynamic order parameter \cite{bkc}
\begin{equation}
A_{loop} =\oint y.df,
\end{equation}
which depends upon $F$ and $\nu$. 
If  $y(f)$ is less than $5$, we call the system to be in the zipped state, 
where as if $y(f) > 5$, it is in the unzipped state \cite{arxiv}.
At high $\nu$, it is evident  that for small $F$, dsDNA remains in the 
zipped state (Fig. 2a), whereas at high $F$, it is in the unzipped state 
(Fig. 2b), irrespective of initial conformations.  Moreover,  the path of $y(f)$ 
for the force 0 to $F$ is different from that of the path for  
$F$ to 0, which constitutes a hysteresis loop. A decrease in $\nu$ leads to 
a bigger path of the hysteresis loop \cite{sm}. 
Depending on the amplitude, the system starts from the zipped conformation as 
shown in Fig. 2a (or open conformations  shown in Fig. 2b) and then gradually 
approaches the open state (or the zipped state) and back to the 
initial state. 

\begin{figure}[t]
\includegraphics[width=3.25in]{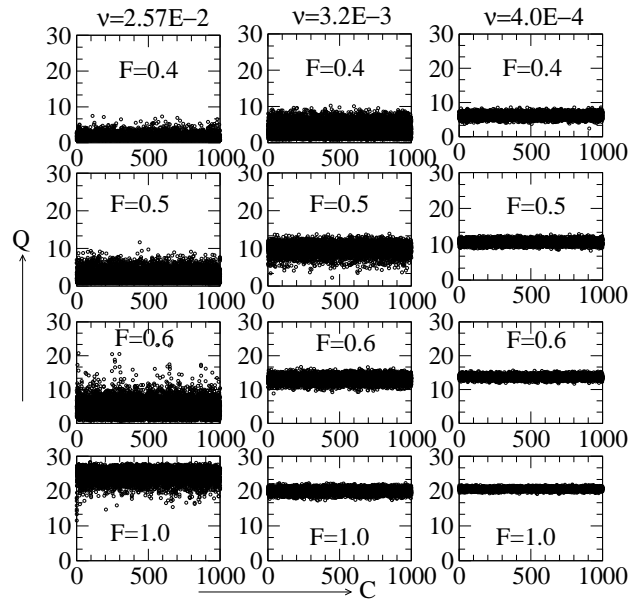}
\caption{The time sequence of $Q$ for different $\nu$ and $F$.}
\end{figure}

One may note that even though $f$ decreases from $F$ to 0 (Fig. 2a), 
$y(f)$ increases and there is some lag, after which it decreases.
Recall that the relaxation time is much higher compare to the time 
spent at each interval of $\Delta f$. Therefore, an increase in $y(f)$ with 
decreasing $f$ indicates that the system gets more time to relax. As 
a result $y(f)$ approaches a path, which is close to the equilibrium. 
Once the system gets enough time, the lag disappears. A similar lag is  
expected, when the system starts from the open state at high $\nu$. 
However, in this case 
as $\nu$ decreases, $y(f)$ decreases with increasing $f$ 
(Fig. 2b). In both cases , whether dsDNA starts from the zipped or open state, 
as  $\nu \rightarrow 0$, the system approaches the equilibrium 
$f-y$ curve and $A_{loop}$ vanishes \cite{sm}. 
Moreover,  at high $\nu$, $A_{loop}$ also vanishes (Fig. 2a \& 2b), but the 
system goes away from the equilibrium.

The other dynamic order parameter
\begin{equation}
Q = \frac{1}{\tau} \oint y(t) dt,
\end{equation}  
studied in the context of magnetic systems \cite{bkc}, has recently been applied
to obtain
the the force-frequency diagram of DNA hairpin \cite{arxiv}. 
In Fig. 3, we plot $Q$ with cycles for different $\nu$ and $F$.  
The distribution shows that the path remains in the zipped state or in the 
open state or in the dynamic (hysteretic) state, depending on $F$ and $\nu$. 
In contrast to the DNA hairpin, which shows the coexistence of different 
states, a dsDNA shows a continuous transition from the zipped state to the 
new dynamic state as the frequency decreases.

We now focus our study on the scaling of the area of the hysteresis loop. 
In Fig. 4a, we have plotted $A_{loop}$ as a function of $(F\nu)^{0.5}$. For 
low $\nu$, we observe that all plots for different $F$ collapse 
on a straight line. This gives the value of $\alpha = 0.5 =\beta$. 
At high $\nu$, depending on the amplitude, the system remains 
either in the zipped state (low $F$) or in the open state (high $F$).
In contrast 
to the spin system, where the average applied field is zero over a cycle, in 
this case, the average applied force is finite over a cycle because the two states 
are asymmetric. In fact, at low $F$, we find that  
$A_{loop}$ scales as $\nu^{-1}(F-f_c)^{2.0 \pm 0.1}$, where $f_c$ is 
the equilibrium critical force at that temperature. The proposed scaling is 
consistent with the  mean field values for a time dependent
hysteretic response to periodic force in case of the isotropic spin \cite{dd}
and found to be independent of length \cite{rakesh}. 

In single-molecule experiments, measurements have been done at 
non-equilibrium conditions. It is possible  to infer the equilibrium 
properties of the system from these data to have a better understanding of the
system. In order to do so, generally measurements have been taken in 
the quasi static limit \cite{prentiss} so that the techniques involved in 
thermodynamics can be employed. In recent 
years, there has been considerable work to extract equilibrium properties from 
the non-equilibrium data {\it e.g.} Jarzynski equality, which relates 
the free energy differences between two equilibrium states through 
non-equilibrium processes \cite{jarzynski}. In another approach 
a dominant reaction pathway algorithm is developed to compute the most probable 
reaction pathways between two equilibrium states \cite{orland}. 

\begin{figure}[t]
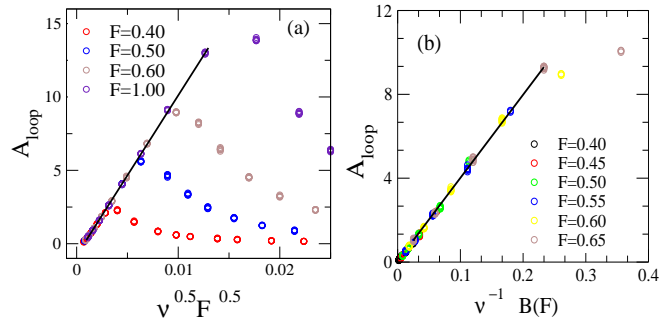

\includegraphics[width=1.7in]{Fig4a.eps}\hspace{0.05in}\includegraphics[width=1.65in,height=1.65in]{Fig4b.eps}
\caption{The scaling of the loop area of hysteresis ($A_{loop}$) with respect to  
$\nu ^{0.5} F^{0.5}$  in the low frequency limit (a) and  with respect 
to $\nu ^{-1}B(F)$ in the high frequency limit. Here, $B(F) \sim (F-{f_c})^{2 \pm 0.1}$ (b).}
\end{figure}

Here, we use the work theorem to derive the equilibrium path between the two 
states \cite{ps11}. Instead of repeating the force cycle $C$ times, we now 
randomly choose $C$ initial conformations, which belong to equilibrium 
conformations at that $T$ and $f$ (=0).
 We follow a similar protocol as described above
to reach the final state (f=F) from the initial state (f=0). No attempt is made to achieve 
equilibrium during this process. The total work performed on the system going 
from the zipped state to open state (forward path) is 
$w_{m_s}=-{\Delta f} \sum_{i=1}^{m_s} y_i$.
When the applied force decreases (backward path) from F to 0, we start with
 $C$ initial conformations, which belong to equilibrium
conformations at that $T$ and $f$ (=F). The work done by the system
 from the open state to the zipped state can be written as  
$w_{1}={\Delta f} \sum_{i=m_s}^{1} y_i$.
The equilibrium distance $y_k$ for the force $f_k$ for the forward path can 
be obtained by assigning the weight $exp(-\beta w_k)$ to all forward $C$ 
paths \cite{ps11} at that instant $k$, which can be written as 
\begin{equation}
y_k=\frac{\sum _{i}^{C} y_k exp(-\beta w_k)}{\sum_{i}^{C}exp(-\beta w_k)}.
\end{equation}
Similarly, $y_k$ for the reverse path can also be obtained.

In Fig.5a, we have plotted the simple average of extension  over many 
($C=1000$) forward paths as well as backward paths ($n=10^4$ LD time steps). 
One can see the existence of hysteresis. In this figure, the solid line 
corresponds
to the equilibrium path, which is the same, whether we start from the zipped 
or open conformation. 
We have used $2\times10^9$ time
steps out of which the first $ 5\times 10^8$ steps have not been taken in the
averaging.  The results are averaged over many trajectories, which are almost
the same within the standard deviation. The weighted average of $y(f)$ 
for the forward and the backward paths obtained from Eq. 3 have also been 
depicted in this plot. One can see from these plots that the weighted 
average even for
$n=10^4$ LD steps is quite close to the equilibrium value ($1.5\times10^9$ LD 
steps). We further note that the weighted average of the backward path almost 
overlaps with the equilibrium path. This  may be because of the fact that in 
a reverse path, two strands of DNA are in the open state and the system can 
access more configurational space. This gives the higher probability of 
choosing  rare conformations, which have dominant contributions in Eq. 3.
It may be noted that the underlying assumption behind the  work 
theorem relies on the fact that the initial state of the system should be 
in the thermal equilibrium. Whereas for the scaling, the system 
needs not be in the equilibrium, but in the steady state. Moreover, 
scaling involves frequency, whereas the equilibrium path obtained from 
the work theorem is independent of frequency.

\begin{figure}[t]
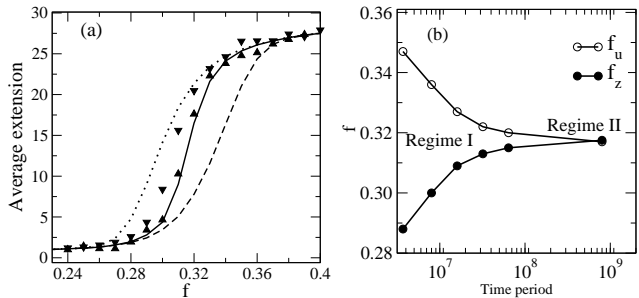

\includegraphics[width=1.7in]{Fig5a.eps}\hspace{0.05in}\includegraphics[width=1.55in]{Fig5b.eps}
\caption{(a) The variation of the average extension as a function of 
force $f$. Dotted and dashed lines correspond to simple average of 
backward and forward paths, respectively. The weighted average of backward 
and forward paths are shown by up and down triangles, respectively.
The solid line represents the equilibrium $f-y$ curve. (b) Simulated values of 
unzipping force ($f_u$) and rezipping force ($f_z$) as a function of time.
Regime I corresponds to the non-equilibrium, where hysteresis has been 
seen \cite{hatch}. Regime II corresponds to the equilibrium, where equilibrium
response of DNA has been studied \cite{prentiss, woodside}.}
\end{figure}

In another study, Chattopadhyay and Marenduzzo \cite{chatto} studied the
dynamics of a polymer chain, whose ends are anchored. An oscillatory force 
was applied at the intermediate bead. 
They also observed hysteresis for the flexible polymer chain. However, 
they showed a crossover from a periodic 
limit cycle (hysteresis) to an aperiodic dynamics as the polymer gets stiffer.
Since the unzipping experiments \cite{prentiss, hatch} usually are performed 
on a long chain (few Kbps) much greater than the persistence length 
of DNA, their model studies \cite{chatto} also imply the existence of 
hysteresis under periodic force \cite{persistense}.

Are the dynamic transition and the scaling proposed here observable in single
molecule experiments? To answer this, in Fig. 5b, we show how the system approaches
the equilibrium (regime II) from the non-equilibrium (regime I). This is
in accordance with experiment followed by simulation \cite{hatch}. For a
two state model, time needed to cross the energy barrier $\Delta E
(10-20 k_BT)$ depending upon the length and sequence of DNA lies in between
4s to 15 min \cite{cocco}. The equilibrium response
of DNA unzipping (regime II) has been studied in experiments belong to this time
scale \cite{prentiss, woodside} as we  obtained in our simulation, but in
($\mu s$) range. There is a mismatch of the time scale because of the
coarse grained description of the model. One of the possible ways to check the 
feasibility of the experiment from our simulation is to compare the 
ratio of time needed for the equilibrium (shown in Fig.2 by turquoise color) 
and the non-equilibrium regime (say black in Fig. 2). From our simulation,
this ratio turns out to be $\sim$ 1000. If the experimental equilibrium time 
is 900 seconds
\cite{prentiss} then the lower limit of time is  900/1000 $\sim$ 1 second.
Hence, by manipulating the amplitude and the frequency in the intermediate time
scale (1s-15 minutes), it is possible to perform experiments where the 
dynamical transition may take place.

In conclusion, we have studied the response of a periodic force on  DNA
unzipping. We showed the existence of a dynamic transition, where by varying the
frequency of the applied force, a dsDNA can go from the zipped/unzipped state 
to a new dynamic (hysteretic) state. The area of the hysteresis loop found to
scale with the same exponents as of spin systems. The scaling exponents are 
found to be quite robust and independent of length \cite{rakesh} and 
friction coefficient \cite{sm}. We also showed that by using the work theorem,  
it is possible 
to extract the equilibrium properties of the system from the non-equilibrium 
data, which have potential application in single molecule force measurements. 
At this stage, additional theoretical and numerical investigations are 
needed to establish a connection between dynamical transition in the spin 
systems and a polymer under periodic force. Since, the role of hysteresis in
biological processes remains unexplored territory, our work calls for
further experiments on periodically driven DNA to explore such hitherto 
unknown dynamical phase transitions and related scaling.

\newpage
We thank S. M. Bhattacharjee, Deepak Dhar, Sriram Ramaswamy and Madan Rao for many helpful
discussions on the subject. We acknowledge the financial supports from the DST and CSIR, 
India. Generous computer supports from MPIPKS Dresden are  gratefully acknowledged by
authors.

\newpage
{\bf Supplementary material} \\
Here, we briefly describe the simulation details adopted in this study.
The model includes a string of beads, each bead represents a base associated with sugar and  phosphate groups.
We consider a sequence in such a way that the first half of the chain is
complementary to the other half. This gives the possibility of the formation
of a dsdNA at low temperature \cite{km}. The energy of the model system,
we adopted, is defined as
\begin{equation}
E = \sum_{i=1}^{N-1}k(d_{i,i+1}-d_0)^2+\sum_{i=1}^{N-2}\sum_{j>i+1}^{N}4(\frac{B}{d_{i,j}^{12}}-\frac{A_{ij}}{d_{i,j}^6}),
\label{eq1}
\end{equation}
where $N(=32)$ is the number of beads. The distance between beads $d_{ij}$, is
defined as $|\vec r_i-\vec r_j|$, where $\vec r_i$ and $\vec r_j$ denote
the position of bead $i$ and $j$, respectively. In the Hamiltonian (Eq. \ref{eq1}), we use dimensionless distances and energy parameters. The
harmonic (first) term with spring constant $k$ (=100) couples the adjacent beads along the chain. The remaining terms correspond to Lennard-Jones (LJ) potential.The  first term of LJ potential takes care of the ``excluded volume effect",
where we set  $B = 1$. We assign the base pairing interaction $A_{ij} = 1$
for native contacts and 0 for non-native ones  \cite{mishra}. This choice
corresponds to the Go model \cite{go}. By native, we mean that the first base
forms pair with the $N^{th}$ (last one) base only and second base with
$(N-1)^{th}$ base and so on as shown in  Fig.1. The parameter $d_0 (=1.12)$
corresponds to the equilibrium distance in the harmonic potential,
which is close to the equilibrium position of the average L-J potential.
In the Hamiltonian (Eq. \ref{eq1}), we use dimensionless distances and energy
parameters. We obtained the dynamics
by using the following Langevin
equation \cite{Allen,Smith}
\begin{equation}
m\frac{d^2r_i}{dt^2} = -{\zeta}\frac{dr_i}{dt}+ F_c+ \Gamma
\end{equation}
where $m(=1)$ and $\zeta(=0.4)$ are the mass of a bead and friction
coefficient, respectively. Here, $F_c$ is defined as $-\frac{dE}{dr}$ and
the random force $\Gamma$ is a white noise \cite{Smith} i.e.,
$<{\Gamma(t)\Gamma(t')}>=2\zeta T\delta(t-t')$.
This keeps temperature constant throughout the
simulation. The equation of motion is integrated using
6$^{th}$ order predictor
corrector algorithm with a time step $\delta t = 0.025$ \cite{mishra}.

Following relations may be used to convert dimensionless units to
real units:  $T= \frac{k_B T^*}{\epsilon}$,
 $t= (\frac{\epsilon}{m {\sigma}^2})^{1/2} t^*$, $r=\frac{r^*}{\sigma}$ \cite{Allen}, where
$T^*$, $t^*$, $r^*$ and $\epsilon$ are temperature, time, distance and energy
in real units. $\sigma$ is the inter particle distance at which potential goes
to zero. For example, if we set effective base pairing energy
$\epsilon \sim 0.1$ eV, we get
$T_m^* \approx 85$ $^0$C, which corresponds to $T_m = 0.23$ in the
reduced unit as reported in Ref \cite{mishra}. It is consistent with the
one obtained by OligoCalc \cite{Kibbe} for the same sequence. Similarly,
by setting the  average molecular  weight of each bead $\sim 308$ g/mol and 
$\sigma=5.17$ $\AA$, we get $t \approx 3.0$ $t^*$ ps. However, this conversion 
is not valid for the entire range of force and temperature \cite{arxiv}.

\section{Dependence of loop area on the applied frequency and amplitude 
of force }
The area under the loop is the  measure of the energy dissipated over a cycle.
At fixed $F$, as
$\nu$ decreases, $A_{loop}$ increases (Fig. 6a).
At some value of $\nu$, $A_{loop}$ is found to be maximum and then it
approaches to zero.
In Fig. 6b, we have depicted the plots of $A_{loop}$ with $F$ for different $\nu$.
In this case also, $A_{loop}$ increases with $F$ and
above a certain amplitude, it starts decreasing. However, one can observe
here that $A_{loop}$ is the slow variant of $F$ and never approaches to zero
for all $F$. 
\begin{figure}[h]
\includegraphics[width=3.4in]{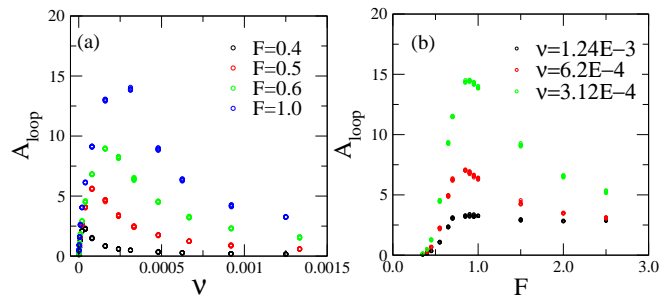}
\caption{Variation of loop area $A_{loop}$  (a) with $\nu$ at different $F$,
(b) with $F$ at different $\nu$.}
\end{figure}

\section{Dependence of scaling exponents on friction coefficient $\zeta$}
We have checked our results for different values of $\zeta$. Here, we have
shown results for 0.4 and 1.2. We found that the scaling exponents in high 
(Fig. 7) and low (Fig. 8) frequency regime, are independent of friction 
coefficient.
\begin{figure}[h]
\includegraphics[width=3.4in]{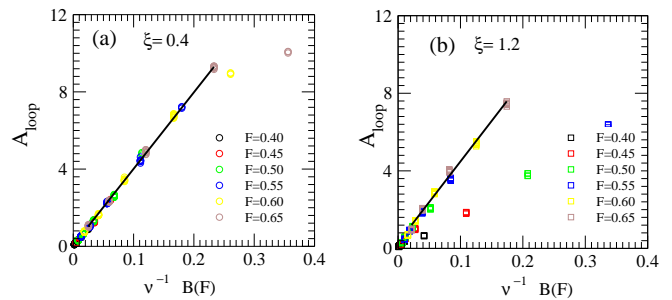}
\caption{Loop area scales as $\nu^{-1}B(F)$ at high frequencies.
 Here, $B(F)\sim(F-f_c)^{2 \pm 0.1}$.}
\end{figure}
\begin{figure}[h]
\includegraphics[width=3.4in]{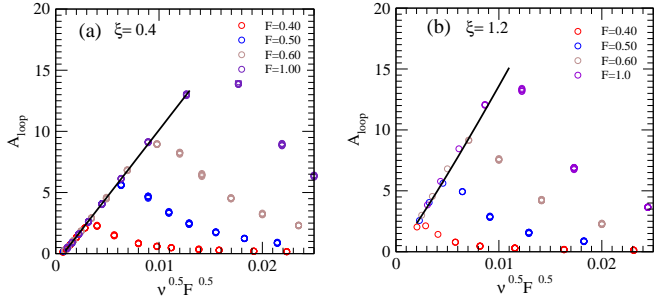}
\caption{Loop area scales as $\nu^{0.5}F^{0.5}$ at low frequencies.}
\end{figure}


\begin{thebibliography}{99}

\bibitem{Wartel} R. M. Wartell and A. S. Benight, Physics Reports {\bf 126},
     67 (1985).
\bibitem{Bockelmann} U. Bockelmann, B. Essevaz-Roulet, and F. Heslot, Phys.
     Rev. Lett. {\bf 79}, 4489 (1997).
\bibitem{prentiss} C. Danilowicz et al., Phys. Rev. Lett. {\bf 93}, 078101 (2004).
\bibitem{bhat99} S. M. Bhattacharjee, J. Phys. A {\bf 33}, L423 (2000).
\bibitem{nelson} D. K. Lubensky and D. R. Nelson, Phys. Rev. Lett. {\bf 85},
     1572 (2000).
\bibitem{Mukamel} Y. Kafri, D. Mukamel, and L. Peliti, Phys. Rev. Lett. {\bf 85},
     4988 (2000).
\bibitem{maren2k2} D. Marenduzzo, S. M. Bhattacharjee, A. Maritan, E. Or-
     landini, and F. Seno, Phys. Rev. Lett. {\bf 88}, 028102 (2002).
     {\bf 268}, 319 (2007).
\bibitem{dnab} I. Donmez and S. S. Patel, Nucl. Acids Res. {\bf 34}, 4216
     (2006).
\bibitem{pcra} S. S. Velankar, P. Soultanas, M. S. Dillingham, H. S.
     Subramanya, and D. B. Wigley, Cell {\bf 97}, 75 (1999).
\bibitem{nphii} M. E. Fairman-Williams and E. Jankowsky, J. Mol. Biol
     {\bf 415}, 819 (2012).
\bibitem{johnson} D. S. Johnson et al Cell {\bf 129} 1299 (2007).
\bibitem{basu} A Basu, A J Schoeffler, J M Berger and  Z Bryant
Nat. Struc. \& Mol. Bio. {\bf 19}, 538 (2012).
\bibitem{Fili} N. Fili et al. NAR {\bf 38} 4448 92010).
\bibitem{Szymczak} P Szymczak and H Janovjak, J. Mol.Biol. {\bf 390}
443 (2009).
\bibitem{kumarphys} S. Kumar and M. S. Li, Phys. Rep. {\bf 486}, 1 (2010).
\bibitem{marko} J. F. Marko and E. Siggia, Macromolecules {\bf 28}, 8759
     (1995).
\bibitem{kumar_prl} S. Kumar {\it et al.},
     Phys. Rev. Lett. {\bf 98}, 128101 (2007).
\bibitem{mishra} G. Mishra, D. Giri, M. S. Li, and S. Kumar, J. Chem.
     Phys {\bf 135}, 035102 (2011).
\bibitem{hatch} K. Hatch, C. Danilowicz, V. Coljee, and M. Prentiss,
     Phys. Rev E. {\bf 75}, 051908 (2007).
\bibitem{kapri} R. Kapri, arXiv: 1201.3709 (2012).
\bibitem{madan1} M. Rao and R. Pandit, Phys. Rev. B {\bf 43}, 3373 (1991).
\bibitem{madan2} M. Rao, H. R. Krishnamurthy, and R. Pandit, Phys. Rev.
     B {\bf 42}, 856 (1990).
\bibitem{dd} D. Dhar and P. Thomas, J. Phys. A {\bf 25}, 4967 (1992).
\bibitem{bkc} B. K. Chakrabarti and M. Acharyya, Rev. Mod. Phys.
     {\bf 71}, 847 (1999).
\bibitem{ps11} P. Sadhukhan and S. M. Bhattacharjee, J. Phys. A. {\bf 43},
     245001 (2010).
\bibitem{sm} See supplementary material for detail.
\bibitem{arxiv} G. Mishra, P. Sadhukhan, S. M. Bhattacharjee, and
     S. Kumar, arxiv: 1204.2913 (2012).
\bibitem{text1} For T = 0.1, critical force $f_c$ was found to be $\sim 0.32$
 \cite{mishra}.
\bibitem{ru} Following relations may be used to convert dimensionless units to
real units:  $T= \frac{k_B T^*}{\epsilon}$,
 $t= (\frac{\epsilon}{m {\sigma}^2})^{1/2} t^*$, $r=\frac{r^*}{\sigma}$ \cite{Allen}, where
$T^*$, $t^*$, $r^*$ and $\epsilon$ are temperature, time, distance and energy
in real units. $\sigma$ is the inter particle distance at which potential goes
to zero. For example, if we set effective base pairing energy
$\epsilon \sim 0.1$ eV, we get
$T_m^* \approx 85$ $^0$C, which corresponds to $T_m = 0.23$ in the
reduced unit as reported in Ref \cite{mishra}. It is consistent with the
one obtained by OligoCalc \cite{Kibbe} for the same sequence. Similarly by setting, average molecular
 weight of each bead $\sim 308$ g/mol and $\sigma=5.17$ $\AA$, we get 
$t \approx 3.0$ $t^*$ ps. However, this conversion does not work in the 
entire range of $f$ and $T$ \cite{arxiv}.
\bibitem{Allen} M. P. Allen and D. J. Tildesley, Computer simulations of
     liquids (Oxford Science, 1987).
\bibitem{Kibbe}W. A. Kibbe, Nucl. Acids Res. {\bf 35}, W43 (2007).
\bibitem{text2} We have analyzed 80 conformations, but in Fig. 2, only
     10 conformations are shown.
\bibitem{rakesh} R. K. Mishra {\it et al.} to be published.
\bibitem{jarzynski} C. Jarzynski, Phys. Rev. Lett. {\bf 78}, 2690 (1997).
\bibitem{orland} P. Faccioli, A. Lonardi, and H. Orland, J. Chem. Phys.
     {\bf 133}, 045104 (2010).
\bibitem{chatto} A. K. Chattopadhyay and D. Marenduzzo, Phys. Rev. Lett.
 {\bf 98}, 088101 (2007).
\bibitem{persistense} Even in a realistic simulation which includes persistence 
length of dsDNA and ssDNA as well as helical structure in its description, 
one  would expect hysteresis curve with no major change in the time scale 
(emerging from large energy gap) as reduction in entropy due to persistence 
length will be roughly compensated by the mobility DNA.
\bibitem{cocco}S. Cocco, Eur. Phys. J. E {\bf 10}, 153 (2003).
\bibitem{woodside} M. T. Woodside {\it et al.}, PNAS, {\bf 103}, 6190 (2006).
\end{thebibliography}

\begin{thebibliography}{99}

\bibitem{km} S. Kumar and G. Mishra, Phys. Rev. E {\bf 78}, 011907
(2008).
\bibitem{mishra} G. Mishra, D. Giri, M. S. Li, and S. Kumar, J. Chem.
     Phys {\bf 135}, 035102 (2011).
\bibitem{go} N. Go and H. Abe, Biopolymers {\bf 20}, 991 (1981).
\bibitem{Allen} M. P. Allen and D. J. Tildesley, Computer simulations of
     liquids (Oxford Science, 1987).
\bibitem{Smith} D. Frenkel and B. Smit, Understanding molecular simulation (Academic Press, London, 2002).

\bibitem{Kibbe}W. A. Kibbe, Nucl. Acids Res. {\bf 35}, W43 (2007).
\bibitem{arxiv} G. Mishra, P. Sadhukhan, S. M. Bhattacharjee, and
     S. Kumar, arxiv: 1204.2913 (2012).
     \end{thebibliography}
\end{document}